\documentstyle[12pt,moriond,psfig]{article}

 1

\def\HH{${\rm {H_2}}\,\,$}
\def\lya{Ly$\alpha$~}

\def\erg{{\rm erg}}
\def\sr{{\rm sr}}

\def\cm{{\rm cm}}

\def\gs{\mathrel{\raise1.16pt\hbox{$>$}\kern-7.0pt
\lower3.06pt\hbox{{$\scriptstyle \sim$}}}}
\def\ls{\mathrel{\raise1.16pt\hbox{$<$}\kern-7.0pt
\lower3.06pt\hbox{{$\scriptstyle \sim$}}}}
\def\gtsima{$\; \buildrel > \over \sim \;$}
\def\ltsima{$\; \buildrel < \over \sim \;$}
\def\prosima{$\; \buildrel \propto \over \sim \;$}
\def\gsim{\lower.5ex\hbox{\gtsima}}
\def\lsim{\lower.5ex\hbox{\ltsima}}
\def\simgt{\lower.5ex\hbox{\gtsima}}
\def\simlt{\lower.5ex\hbox{\ltsima}}
\def\simpr{\lower.5ex\hbox{\prosima}}

\def\pp{\noindent\parshape 2 0truecm 17truecm 2truecm 15truecm}
\def\rf#1;#2;#3;#4 {\par\pp#1, #2, #3, #4. \par}

\def\pr{\ref@jnl{Phys.Rev}}

\def\href#1;#2 {{\bf #1} : {\em #2}}

\def\beq#1{\begin{equation}\label{#1}}
\def\eeq{\end{equation}}
\def\beqa#1{\begin{eqnarray}\label{#1}}
\def\eeqa{\end{eqnarray}}

\def\tento#1{\times 10^{#1}}

\def\s{{\rm \ s}}
\def\sr{{\rm \ sr}}

\def\erg{{\rm \ erg}}
\def\cm{{\rm \ cm}}

\def\Hz{{\rm \ Hz}}

\def\HH{H$_2$ }
\def\H2p{H$_2^+$ }

\def\mH2p{H_2^+}

\begin{document}
\heading{Pop III objects and their Relationship with Dwarf Galaxies}

\author{A. Ferrara  $^{1}$, B. Ciardi $^{2}$} 
%{$^{1}$ Joint Institute for Laboratory Astrophysics, Boulder, USA} 
{$^{1}$ Osservatorio Astrofisico di Arcetri, Firenze, Italy} 
{$^{2}$ Dipartimento di Astronomia, Universit\'a di Firenze, Italy}

\begin{moriondabstract}
In this paper we briefly review some of the most recent developments
concerning the formation of the first luminous objects in the universe,
to which we refer to as Pop III objects, and their observable
effects. In particular, we try to critically discuss the possible
connection between them and dwarf spheroidal galaxies. We come to the
conclusion that, if the properties of Pop IIIs are correctly predicted
by hierarchical cosmological models, their identification with present
day dwarf galaxies presents several difficulties.
\end{moriondabstract}

\section{The First Luminous Aggregations in the Universe}

Determining how structure formed in the early universe, and how
the first luminous objects (hereafter Pop III) form, appears to
be one of the most challenging and interesting problems in 
physical cosmology. 
Current models of cosmic structure formation based on CDM scenarios
predict that the first collapsed, luminous (hereafter Pop~III) objects 
should form at redshift $z\approx 30$ and have a total mass 
$M \approx 10^6 M_\odot$ or baryonic mass $ M_b \approx 10^5 M_\odot$
(\cite{CR}, \cite{HRL}, \cite{TSR}, \cite{CF}). 
This conclusion is reached by requiring
that the cooling time of the gas is shorter than the Hubble time 
at the formation epoch. In a plasma of primordial composition 
the only efficient coolant in the temperature range $T\le 10^4$~K, the 
typical virial temperature of Pop~III dark matter halos, is represented 
by H$_2$ molecules, whose abundance increases
from its initial post-recombination relic fractional value ($f_{H2} 
\approx 10^{-6}$) to higher values 
during the halo collapse phase. It is therefore crucial to determine 
the cosmic evolution of such species in the early universe to clarify 
if small structures can continue to collapse according to the postulated 
hierarchical structure growth or if, lacking a cooling source, the mass 
build-up sequence comes to a temporary halt.

During the collapse, if fragmentation occurs, a stellar cluster is likely 
to be formed with a stellar mass, $M_*$, which depends on the (unknown) 
details of the star formation process. 
As an order of magnitude estimate, we can calculate the mass of stars formed
in a free-fall time using the following simple formula (\cite{AF}):
\begin{equation}
\label{Ms}
M_*\approx{\Omega_b f_b \over\tau }M\simeq
300 \Omega_{b,5} f_b M_6 ~~M_\odot
\end{equation}
The baryon density parameter is $\Omega_{b}=0.05~\Omega_{b,5}$, of
which a fraction $f_b$ is able to cool
and become available to form stars. The total mass of the 
object is $M=10^6 M_6~M_\odot$; finally, $\tau^{-1}=0.6 \%$ is the 
star formation efficiency, calibrated on the Milky Way.
Clearly, large uncertainties affect the three free parameters entering
equation (\ref{Ms}), and in particular $f_b$. Generally speaking, the 
first objects should have a baryonic mass of about 10$^5~M_\odot$ 
and form about 100-1000 stars during the first episode of star formation.
Thus, given these derived properties, a Pop III object might look 
very similar to a star-forming molecular cloud in the Galaxy, and indeed
they are very likely to be the first molecular, self-gravitating objects
in the universe.  

Some of the formed stars could be more massive than $8~M_\odot$. 
Their number crucially depends on the actual IMF in those conditions,
which, needless to say, is almost totally unknown. However, if we
assume a standard Salpeter-like IMF with a low-mass cutoff at 0.1
$M_\odot$, then one such star is produced for each 56~$M_\odot$ 
of stars formed, or, according to the previous estimate, about 
2-20 massive stars. 
Massive stars are particularly important for two reasons: (i) they
produce ionizing photons, and (ii) end their lives exploding as
supernovae. In turn, photons with energies above 13.6 eV produce ionized 
regions that expand in the surrounding IGM, a cosmological analog 
of galactic
HII regions. However, in addition to hydrogen, if the ionizing 
spectrum is hard enough, similar ionized spheres may occur for 
helium as well. Pop III are thus certainly initiating
the process of reionization of the universe, which was essentially 
neutral since
the recombination epoch at $z\approx 1100$. It is not completely
clear, though, if they can bring this process to 
completion and on which timescale. Also, keeping the universe ionized 
might require additional ionizing sources - maybe larger galaxies or QSOs.    
This uncertainty is due to the fact that the radiation
emitted by star-forming Pop IIIs can photodissociate H$_2$ 
molecules that are necessary to continue the hierarchical collapse
sequence. In this sense, Pop IIIs are a population potentially committing
self-suicide~!
Massive stars also provide mechanical energy input to
the parent galaxy ISM via winds, and most importantly, supernova
explosions. The effects of such energy injection can be dramatic
in low-mass objects (for a detailed discussion see 
\cite{MLF}), possibly   leading to the complete dispersal of
the parent galaxy ISM. In that case the star formation activity is 
subsequently quenched by the lack of gas and only a stellar remnant
made of low-mass stars is left over. Moreover, such pristine supernovae
contribute to the pollution of the IGM with heavy elements, which
are now detected by absorption line experiments toward QSOs. 
The first solid particles in the universe, i.e. dust grains, could 
also be formed in the expanding Pop III SN ejecta: such a possibility 
has recently become clear after the SN1987A event. Finally, 
together   with reionization, reheating of a
fraction of the IGM to temperatures of the order of a million degrees
must occur as a result of shocks produced by explosions. 

In the following  we discuss radiative and mechanical 
feedback due to Pop III objects in more detail. Some of the presented 
results will provide us with some ground to speculate about
the possible connection between Pop III objects and local dwarf galaxies. 

\section{Radiative Feedback Effects}

If massive stars form in Pop~III objects, their photons with $h\nu > 13.6$~eV
create a cosmological HII region in the surrounding IGM.
Its radius, $R_{i}$, can be estimated by solving the following standard 
equation for the evolution of the ionization front:
\begin{equation}
\frac{dR_{i}}{dt} -H R_{i}= \frac{1}{4 \pi n_{H} R_{i}^{2}} \; \left[
S_{i}(0) - \frac{4}{3} \pi R_{i}^{3} n_{H}^{2} \alpha^{(2)} \right];
\label{HIIrad}
\end{equation}
note that ionization equilibrium is implicitly assumed.
$H$ is the Hubble constant, $S_{i}(0)$ is the ionizing photon
rate, $n_{H}= 8\times 10^{-6} \Omega_b h^2 (1+z)^3$~cm$^{-3}$ is the
IGM hydrogen number density
and $\alpha^{(2)}$ is the hydrogen recombination rate to levels $\ge 2$.
First, we note that
since $R_i \ll c/H$, the cosmological expansion term $HR_i$ can be
safely neglected.
In its full form eq.~(\ref{HIIrad}) must be solved numerically \cite{CAF}. 
However, in some limiting cases $R_i$ can be evaluated analytically. 
If steady-state is assumed ($dR_i/dt \simeq 0$), then $R_i$ is approximately
equal to the Str\"omgren radius, $R_{S}=[3 S_{i}(0)/(4 \pi n_{H}^{2}
\alpha^{(2)})]^{1/3}$. In general, $R_{s}$ represents an upper limit
for $R_{i}$, since the ionization front completely fills the time-varying
Str\"omgren radius only at very high redshift, $z \approx 100$. 
For our reference parameters it is:
\begin{equation}
R_{i} \le R_{s} = 0.05 \, \left( {\Omega_b h^2} \right)^{-2/3}
(1+z)_{30}^{-2} S_{47}^{1/3} {\rm ~~kpc},
\label{rion}
\end{equation}
where $S_{47}=S_{i}(0)/(10^{47}$ s$^{-1}$). 
The approximate expression for $S_i(0)$ is $\approx 1.2 \times 10^{48}
f_b M_6$~s$^{-1}$ when a 20\% escape fraction for ionizing photons
is assumed (\cite{CAF}). A detailed comparison (\cite{CAF}) between
$R_{i}$ and $R_{s}$ shows that $R_{s}$ is
typically 1.5 times larger than $R_{i}$ in this redshift range.

As mentioned above, in addition to a cosmological HII region, photons
with 11.26 eV $\le h\nu \le$13.6 eV, create a photodissociated sphere
by exciting the \HH Lyman-Werner bands (LW) which eventually decay into the
continuum (this is the so-called two-step photodissociation process).
The main difference between ionization and dissociation sphere
evolution consists in the fact that at high $z$ there is no efficient mechanism to
re-form the destroyed H$_{2}$, analogous to H recombination. As a
consequence, it is impossible to define a photo-dissociation Str\"omgren
radius. Given a point source that radiates $S_{LW}=\beta S_i(0)$ photons 
per second in the LW bands, an estimate of the maximum
radius of the H$_2$ photodissociated
sphere, $R_{d}$, is the distance at which the photo--dissociation time
becomes longer than the Hubble time:
\begin{eqnarray}\label{rcrit}
R_{d} \simlt 2.5 \, h^{-1/2} (1+z)_{30}^{-3/4} S_{LW,47}^{1/2} \;\; {\rm kpc},
\end{eqnarray}
where $S_{LW,47}=S_{LW}/(10^{47}$ s$^{-1}$).
Eqs.~(\ref{rcrit}) and~(\ref{rion}) show that the photodissociated
region
is larger than the ionized region; however, even if $S_{LW} \ll S_i(0)$
under most conditions $R_d$ cannot be smaller than $R_i$ since inside
the HII region, H$_2$ is destroyed by direct photoionization.

At high ($z\approx 20-30$) redshifts the mean separation between Pop III
objects is larger compared to the photodissociated sphere radius.
In between the dissociated spheres, a "soft" UV background (SUVB)
is established, which is able to  photodissociate intergalactic \HH 
molecules in a time shorter than the Hubble time only  if its intensity 
exceeds the critical value
\begin{equation}
J^{crit}_{LW} = 6.2\tento{-4} J_{21} h  (1+z)_{30}^{3/2},
\end{equation}
where $J_{21}=10^{-21}\erg\s^{-1}\Hz^{-1}\cm^{-2}\sr^{-1}$. 
The SUVB intensity can then be calculated to a good approximation
(for the detailed derivation and discussion see \cite{CAF}) 
in the limit of small volume filling
factor of the photodissociated spheres; overlapping starts to be
important at redshift $\simlt 20$. The calculation of the above
authors includes intergalactic H$_{2}$ attenuation, the effects 
of cosmological expansion and neutral H line absorption.
\begin{figure}
\centerline{\psfig{figure=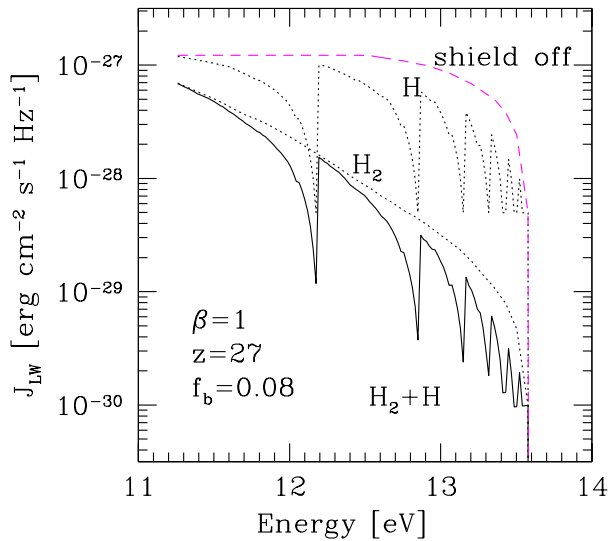,height=10cm}}
\caption{\label{fig1}{Spectrum of the SUVB,
$J_{LW}$, for $z=27$, $\beta=1$ and for baryon cooling efficiency
$f_{b}=0.08$; from the top to the bottom, no shielding (dashed line),
H lines opacity (dotted),
H$_{2}$ lines opacity (dotted) and H$_{2}$ and H lines opacity (solid)
is included.}}
\end{figure}
In Fig.~\ref{fig1} the spectrum of the SUVB at $z=27$
is shown, for the particular value of the baryon cooling efficiency 
$f_{b}=0.08$ (this  fixes the intensity of the radiation field) and 
in which the IGM
LW attenuation is either included or neglected for comparison sake.
As $J_{LW}$ depends linearly on $\beta$, the 
ratio between the flux just below and above the Lyman limit, 
the results for different values
of $\beta$ can be easily obtained by appropriately scaling the plotted curves.
Fig.~\ref{fig1} allows to conclude that the intergalactic H$_{2}$ absorption
cannot be neglected as it decreases the SUVB intensity by more than 
an order of magnitude at energies just
below 13.6 eV, while the attenuation is reduced at lower energies by the
cosmological expansion.
Also, by inspecting Fig.~\ref{fig1} it is possible to compare the relative
effects of hydrogen line and H$_{2}$ LW line absorption
on the SUVB attenuation, for a
typical choice of the parameters. The H lines are optically thick at
their center; this, combined with the effect of the cosmological
expansion, produces the typical sawtooth modulation of the spectrum.
On the other hand, the radiative decay of the excited H atoms produces
Ly$\alpha$ and other Balmer or lower line photons, which are out of the LW
energy range and result in an increase of the flux just
below the Ly$\alpha$ frequency. From Fig.~\ref{fig1} we see that the
molecular line attenuation dominates  the hydrogen one over the entire
range of frequencies.

These results are particularly important to estimate the relevance of 
the so-called "negative feedback" effects (\cite{HRL}). In brief,
the SUVB created by pregalactic objects, could 
penetrate large clouds, and, by suppressing their \HH abundance,
prevent the collapse of the gas. However, the low values of the 
SUVB found at high redshift are well below the threshold required 
($J_{LW}=10^{-24}\erg\s^{-1}\Hz^{-1}\cm^{-2}\sr^{-1}$) for the 
negative feedback (\cite{HRLE}), note that there was an error
in their original paper) on the subsequent 
galaxy formation to be effective. A complete discussion on the
issue of negative feedback can be found in \cite{CAF}.

\section{High-z SNe and the Fate of Pop IIIs}

If stars more massive than $8 M_\odot$ are born in
the Pop III stellar cluster, supernovae will start to explode 
after a few million years.
As discussed above, for a Salpeter IMF we expect 
about $N=2-20$ supernovae to blow off inside a Pop III object. 
The total mechanical luminosity of such a multi-SN explosion, approximated
as a continuous energy injection,  is of the order of  $\epsilon_0 N/t_{_{OB}}
\simeq 0.6-6\times 10^{37}$~erg~s$^{-1}$ (\cite{AF}), where 
$\epsilon_0 \approx 10^{51}$~erg
is the energy of a single supernova explosion and $t_{_{OB}} 
\approx 10^7$~yr is the typical lifetime of a massive star association. 
The point to be appreciated is that
Pop III objects, due to their low mass and binding energy, are very fragile
objects and tend to be {\it blown-away} (\cite{CF}) 
by explosions:
a number $N_c \simeq M_6^{5/3} (1+z)_{30} h^{2/3}$ of supernovae will suffice to
disrupt the object. Since for our reference
stellar cluster $N > N_c$, its residual gas content
will be swept away by the expanding multi-SN driven shock and lost to
the IGM. 
As a result, further star formation will be inhibited and the 
remaining low mass stellar population will continue to evolve passively. 
The gas injected in the IGM is enriched by the heavy elements produced by 
supernova nucleosynthesis
processes. Assuming that the mass in heavy elements averaged on a
Salpeter IMF is $\approx 3 M_\odot$ we estimate that the metallicity 
of the hot gas inside the IGM swept-up shell corresponding to a 
blow-away event driven by $N$ supernovae is
\begin{equation}
\label{z}
Z \simeq 1.7\times 10^{-4} N^{2/5}(1+z)_{30}^{18/5} (\Omega_b h
^2)^{-2/5},
\end{equation}
or $Z \simeq 0.05 N^{2/5} Z_\odot$ at $z\approx 30$ and $(\Omega_b h^2)=
0.05$. This estimate shows that Pop III objects might be responsible for the
origin of polluted regions of relatively high metallicity, althought of rather
small (sub-kpc) size. Hence, the enrichment of the universe might very well 
have been very patchy in its early phases (\cite{GO}).

\begin{figure}
\centerline{\psfig{figure=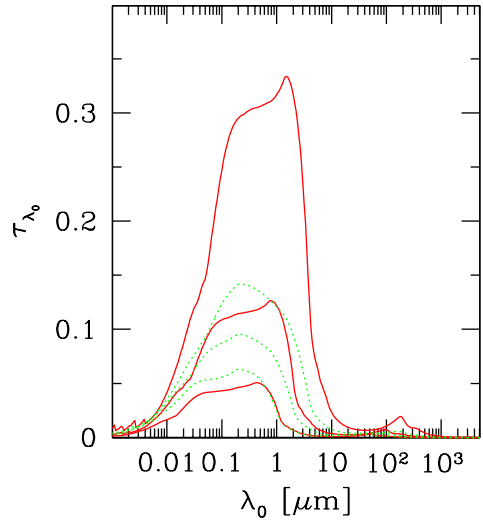,height=10cm}}
\caption{\label{fig2}{Dust opacity as a function of the present
observation wavelength. The curves refer to
different values of the metallicity evolution parameter $q=0$ 
(no evolution, solid lines),$q=1$ (moderate evolution, dashed)
and redshift $z=5,10,20$ from the lowermost to the uppermost 
curves, respectively. For details see \cite{FNSS}.
}}
\end{figure}

Interesting conclusions can also be drawn from the study of dust in the IGM,
and, in particular, in Ly$\alpha$ clouds.
In fact, if Pop III SNe act as metal pollutors for the IGM, 
it is very likely that some dust (which is now firmly established to 
form in SN ejecta) is also
injected at the same time. In a recent paper (\cite{FNSS}) we have 
investigated the possible observational consequences of dust 
in \lya forest clouds. We relate the dust content, $\Omega_d^{Lya}$, 
to the metal evolution of the absorbers and assume that dust is heated 
by the ultraviolet background radiation
and by the CMB.  We find that the dust temperature deviates from the
CMB temperature by at most 10\% at redshift $z=0$.
The \lya cloud dust opacity to redshift $\sim 5$ sources around the
observed wavelength $\lambda_0 \sim 1 \mu$m is $\sim 0.13$, and could
affect observations of the distant universe in that band. The dust 
opacity as a function of redshift and wavelength for two different 
metallicity evolution scenarios is shown in Fig. \ref{fig2}. 
The expected CMB spectral distortions due to reradiation by 
high-$z$ dust in \lya clouds is shown to be $\sim 1.25-10$ times 
smaller than the current COBE upper limit, depending on
the metallicity evolution of the clouds.

\section{Are dSphs Dead Pop III Remnants ?}

The final fate, and consequently the nature of the possible 
local counterparts of Pop III objects, are not yet clear. For
example, stars can become unbound due to dynamical
instabilities following the blow-away or they can steadily evaporate 
from the stellar cluster and be lost in the field; the timescale of 
these processes, though, should be compared with the one for merging.
Nevertheless, it is conceivable that some of these primordial structures
might have survived up to the present epoch. Clearly, the most suitable 
counterpart candidates are dwarf spheroidal galaxies (an extensive review on dSphs 
is given in \cite{FB}), which could be tentatively identified with Pop 
III objects that underwent blow-away but managed to survive merging 
into larger structures. This idea has also been proposed recently by
\cite{MR}. It is tempting, therefore, to explore this hypothesis further
by comparing different properties of the two classes of objects.  
This comparison is schematically shown in Fig. 3, where we consider
seven different parameters: total mass, age, size, stellar mass, star
formation history, metallicity and merging history; in the following we 
discuss them in detail and highlight possible problems and positive
aspects of such identification.
\begin{figure}
\centerline{\psfig{figure=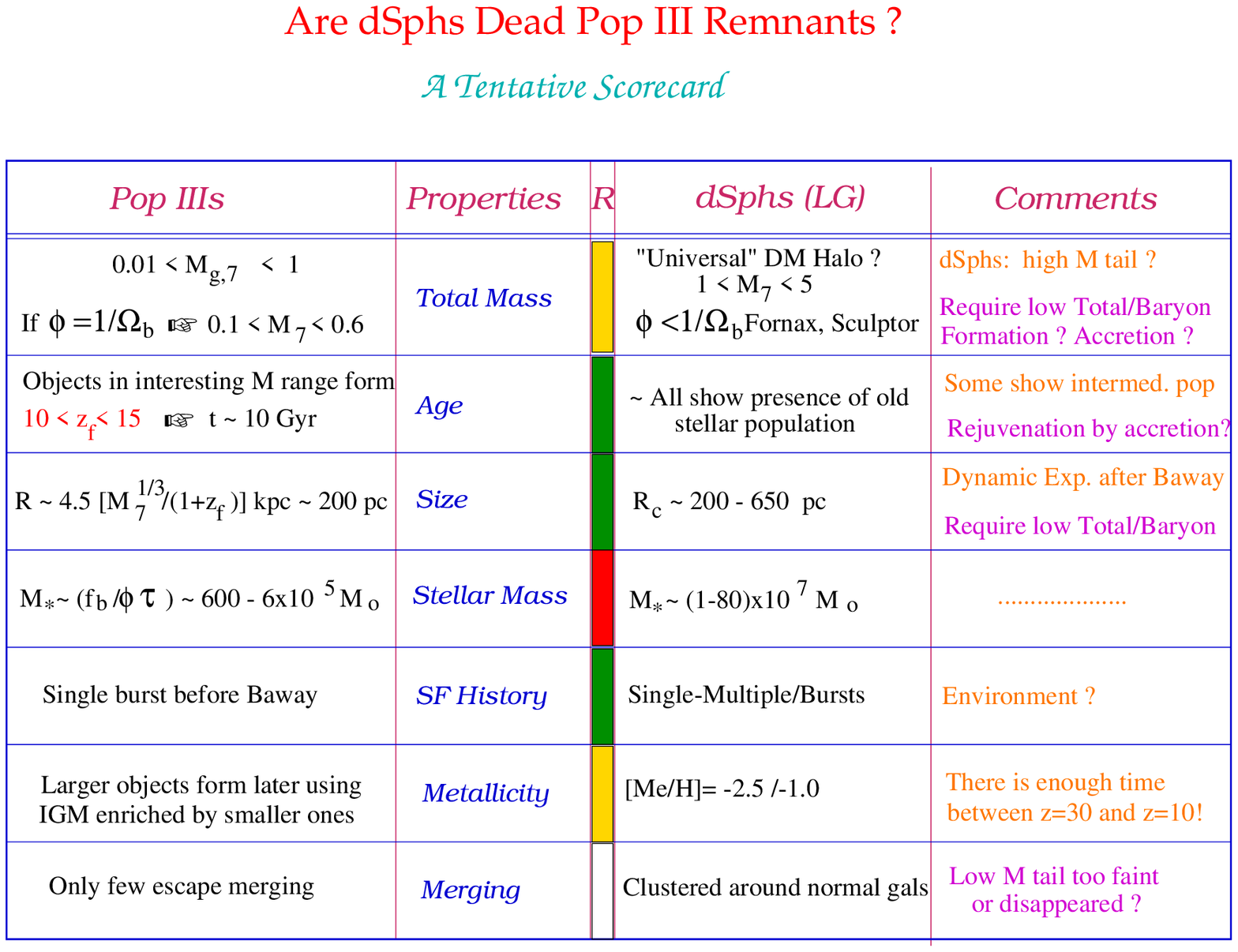,height=10cm}}
\caption{\label{fig3}{Comparison among various properties of
Pop III objects and dwarf spheroidal (dSph) galaxies. 
}}
\end{figure}
The expected initial gas mass (in units of $10^7 M_\odot$) 
of Pop IIIs is $0.01 \simlt M_{g,7} \simlt 0.1$,
where the lower limit is dictated by the cooling timescale condition 
for formation
discussed above and the upper one corresponds to the largest
mass that cannot escape blow-away. 
The last condition is shown in Fig. \ref{fig4}, in the
plane $\phi-M_g$ ($\phi$ is the initial total to baryonic mass ratio
of the object) along with the regimes in which either blow-out (essentially
a wind rather that a complete ejection of the baryons as in the 
blow-away case, see \cite{MLF}) or no mass loss is taking place. 
In addition, in Fig. \ref{fig4} we have plotted the collapse redshift 
as a function of mass in a CDM model with spectrum power index $n=2$;
again, we see that above $z=30$ the cooling time becomes larger than
the Hubble time and hence the collapse of small objects is inhibited. 
If $\phi$ takes the cosmological dark-to-baryonic ratio $\Omega_b^{-1}=16.6$ 
(dashed horizontal line in Fig. \ref{fig4}), then 
the above mass range shift into  $0.1 \simlt M_7 \simlt 0.6$, 
where $M_7=M/10^7 M_\odot$ is the total mass. 
\begin{figure}
\centerline{\psfig{figure=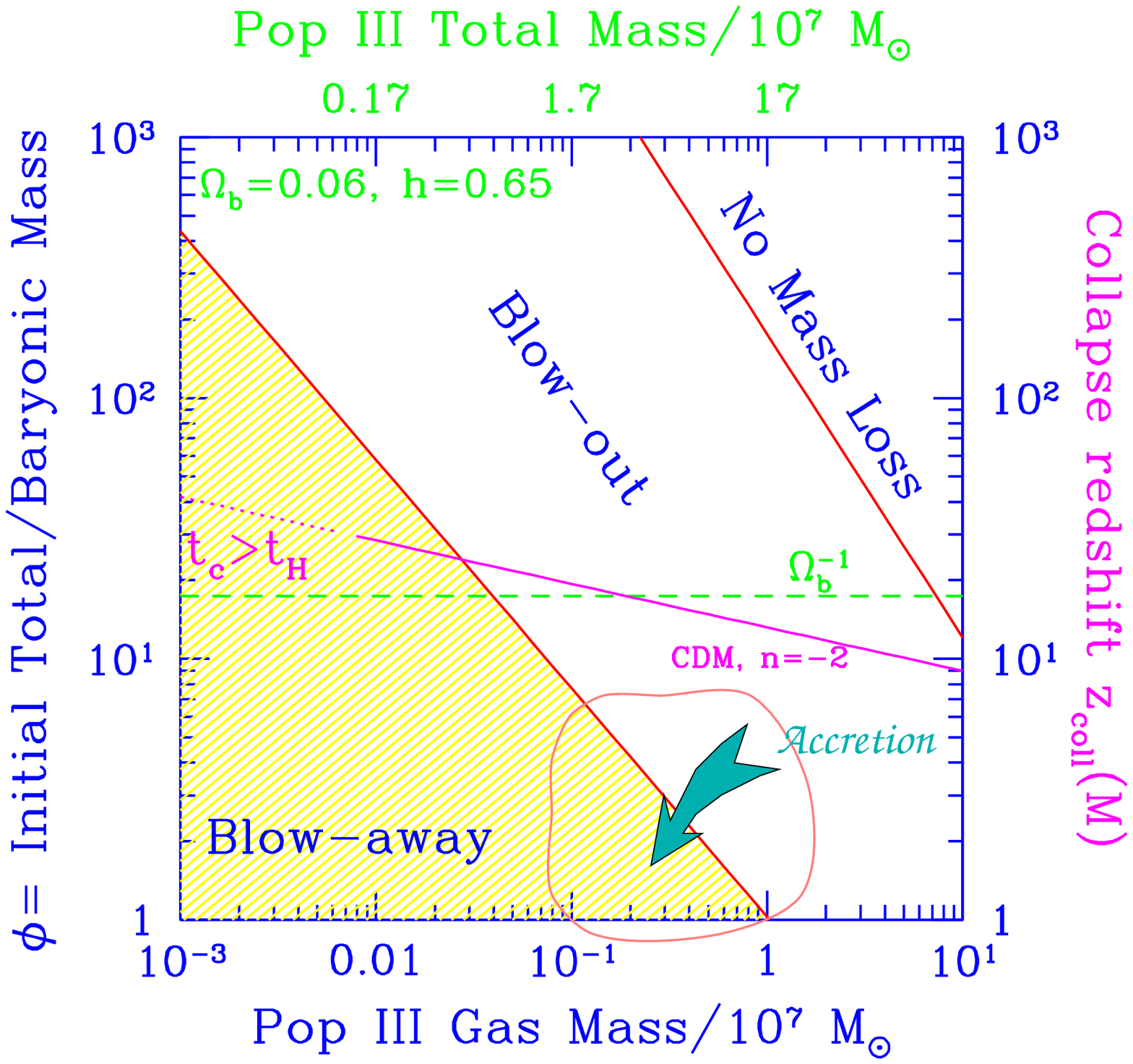,height=10cm}}
\caption{\label{fig4}{Regions in the $\phi-M_g$ plane in which
blow-away, blow-out or mass loss respectively occur in a Pop III object.
Also shown is the collapse redshift of the objects in a CDM model with
$n=2$ as a function of $M_g$; for redshift higher than $z=30$ (dashed line)
no collapse is possible. The horizontal line show the cosmological
value of $\phi=\Omega_b^{-1}$.
}}
\end{figure}
The typical mass of dSph dark matter halos are instead in the range 
$1 \simlt M_7 \simlt 5$. However, for at least two galaxies 
(Fornax and Sculptor) the value of $\phi \ll \Omega_b^{-1}$; this 
could be explained by the
fact that these galaxies have been formed in a gas rich environment
in which baryonic accretion was relatively easy. These galaxies
might have been initially in the blow-out regime, but after an 
outflow phase in which a fraction of the gas is lost, they entered 
the regime in which blow-away was 
unavoidable. We conclude that observed dSphs in principle could be 
associated with the high
mass tail of the Pop III distribution; lower mass objects could be 
simply faint enough to have escaped detection to date. Pop III objects in the 
mass range overlapping dSphs form at redshift $z=10-15$, corresponding
to a present age of about 10 Gyr. Essentially all the dSphs in the Local
Group show the presence of an old stellar population; however, in some of
them (the prototypical example being Carina) the analysis of color-magnitude
diagrams has convincingly demonstrated that at least one subsequent star 
formation episode might have taken place. This represents a serious problem
for the blown-away Pop III scenario, which could be reconciled only
if some rejuvenation by accretion or some other gas input to the system
(returned from low mass stars ?) is postulated. To alleviate the problem, 
it has to be noted that this intermittent star formation activity is 
not explained in detail by any current theoretical model.   
This particular star formation history might suggest, as for the 
low observed values of $\phi$, that dSphs might be Pop III that 
formed/liveed in particularly high density environments.
Sizes of the two classes of object are compatible, being of the order
of tenths of kpc; again, dSphs are found to be slightly larger.
Probably the major difficulty for the identification hypothesis is
represented by the stellar content, which, as shown in
Fig. \ref{fig3}, is about a factor of $\approx 10$ larger for dSphs. 
Implicitly,
this discrepancy has its roots once again in the fact that dSphs
are more baryon-rich than canonical Pop IIIs. 
Finally, the metallicity-mass relation commonly observed in dSphs
can be qualitatively explained in the framework of hierarchical 
structure formation models: in fact, as larger objects form, they 
use IGM previously metal enriched by smaller ones. The time interval required
for the metallicity build-up does not seem to represent a particular 
problem between redshift 30 and 10, when the oldest generation of
stars has to be in place. As mentioned above, some of the Pop IIIs
must be swallowed/merged into larger objects; the fraction of them
that has survived such events can be rather small.  

Clearly, the comparison carried out in Fig. 3 and in its relative
discussion can be seen only as a tentative scorecard and it remains
highly speculative. Nevertheless, it is clear that a one-to-one   
identification between Pop III objects, as theoretical predictions
describe them, and dSph galaxies, as we observe them today, appears
to be difficult to maintain. 

% References listed in alphabetical order ...

%
\vfill
\end{document}